\documentclass[aps,10pt,prl,twocolumn,letterpaper,superscriptaddress]{revtex4-2}

\usepackage{graphicx} % Include figure files
\usepackage{bm} % bold math
\usepackage{amsmath,amssymb}
\usepackage{xcolor}

\definecolor{PRLblue}{rgb}{0.18,0.18,0.57}
\usepackage[colorlinks=true,allcolors=PRLblue]{hyperref} % add hypertext capabilities
\graphicspath{ {./}{figs/}}
\newcommand{\myhash}{\scalebox{0.8}{\raisebox{0.4ex}{\#}}}

\usepackage{fancyhdr}
\begin{document}
%\title{ Antiferromagnetic topological insulator near the metal-insulator transition in MnS$_2$ }
\title{Antiferromagnetic $\mathbb{Z}_2$ topological metal near the metal-insulator transition in MnS$_2$}

\author{Vsevolod Ivanov}
\affiliation{Virginia Tech National Security Institute, Blacksburg, Virginia 24060, USA}
\affiliation{Department of Physics, Virginia Tech, Blacksburg, Virginia 24061, USA}
\author{Xiangang Wan}
\affiliation{National Laboratory of Solid State Microstructures, School of Physics and Collaborative Innovation Center of Advanced Microstructures, Nanjing University, Nanjing, China}
\author{Sergey Y. Savrasov}
\affiliation{Department of Physics, University of California, Davis, CA 95616, USA}

%\author{Vsevolod Ivanov$^{\dag }$, Xiangang Wan$^{\ast }$, Sergey Y. Savrasov%	$^{\dag }$}
%\affiliation{$^{\dag }$Accelerator Technology and Applied Physics Division, Lawrence Berkeley National Laboratory, Berkeley, CA 94720, USA Molecular Foundry,  Lawrence Berkeley National Laboratory, Berkeley, CA 94720, USA}
%\affiliation{$^{\dag \dag}$Department of Physics, University of California, Davis, CA 95616, USA}
%\affiliation{$^{\ast }$National Laboratory of Solid State Microstructures, School of Physics and Collaborative Innovation Center of Advanced Microstructures, Nanjing University, Nanjing, China. }

\begin{abstract}	
Antiferromagnetic (AFM) semiconductor MnS$_2$ possesses both high-spin and low-spin magnetic phases that can be reversibly switched by applying pressure. With increasing pressure, the high-spin state undergoes pressure-induced metalization before transforming into a low-spin configuration, which is then closely followed by a volume collapse and structural transition. We show that the pressure driven band inversion is in fact topological, resulting in an antiferromagnetic $\mathbb{Z}_2$ topological metal (Z2AFTM) phase with a small gap and a Weyl metal phase at higher pressures, both of which precede the spin-state crossover and volume collapse. In the Z2AFTM phase, the magnetic order results in a doubling of the periodic unit cell, and the resulting folding of the Brillouin zone leads to a $\mathbb{Z}_2$ topological invariant protected by the persisting combined time-reversal and half-translation symmetries. Such a topological phase was proposed theoretically by Mong, Essin, and Moore in 2010 for a system with AFM order on a face-centered cubic (FCC) lattice, which until now has not been found in the pool of real materials. MnS$_2$ represents a realization of this original proposal through AFM order on the Mn FCC sublattice. A rich phase diagram of topological and magnetic phases tunable by pressure, establishes MnS$_2$ as a candidate material for exploring magnetic topological phase transitions and for potential applications in AFM spintronics.
\end{abstract}

\maketitle

\section{I. Introduction.}

Non-magnetic topological materials have been extensively explored and classified in readily accessible materials databases \cite{topodatabase1, topodatabase2, topodatabase3}, however, magnetic topological materials are comparatively less studied \cite{topomag1, topomag2}. In fact there are few examples of magnetic topological insulators, most of which are based on the Bi$_2$Se/Te$_3$ family of materials \cite{bi2se3} into which magnetism is introduced by doping \cite{cr-bi2te3}, a magnetic substrate \cite{bi2se3-eus}, or through a heterostructure \cite{mnbi2te4, mnbi4te7, mnbi2te4-review}. New materials hosting these kinds of phases can serve as platforms for studying the complex interplay between magnetism and topology, as well as transitions between different topological phases. In particular, the discovery of new antiferromagnetic (AFM) topological materials, is motivated by their potential applications in the field of AFM spintronics \cite{AFspintronics,2DAFTI}.

\begin{figure}[hb]
    %\captionsetup{justification=raggedright}
    \includegraphics[width=0.7\columnwidth]{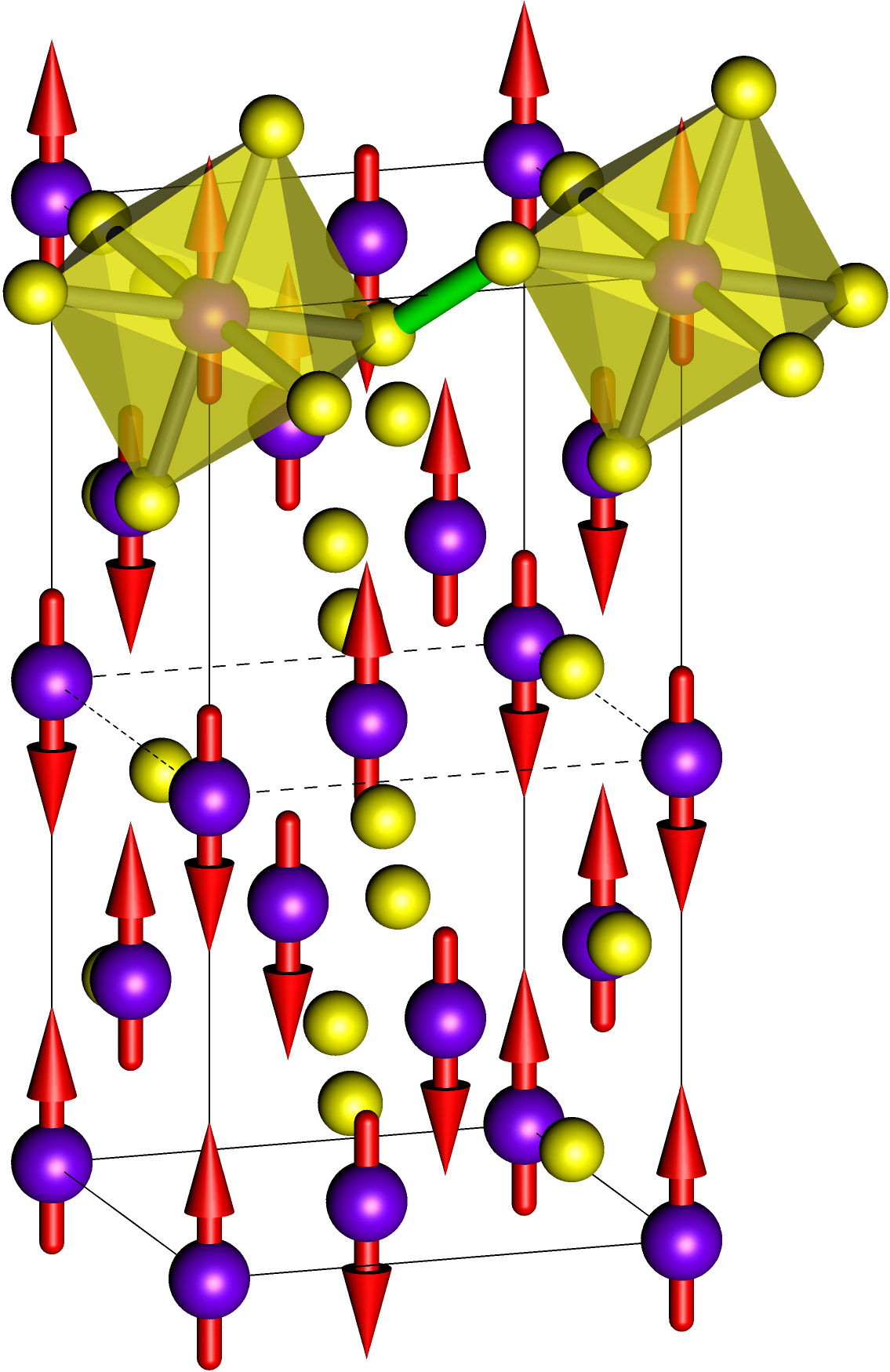}
    \caption{Crystal structure of the magnetic unit cell of MnS$_2$. Manganese atoms (purple) form an FCC sublattice, and sulfur atoms (yellow) surround each Mn atom in an octahedral arrangement. Arrows indicate the magnetic moments on the Mn atoms in the AFM-III phase below $T_N$ = 48K.}
    \label{struct}
\end{figure}

The earliest example of an aniferromagnetic topological insulator (AFTI) \cite{moore-afti} was theoretically proposed shortly after the experimental discovery of the first topological insulator. In this model AFTI, time reversal symmetry $\Theta$ is broken by antiferromagnetism, leading to a doubling of the periodic unit cell. However, the combination of time-reversal and half-translation symmetries of this new unit cell, $\Theta T_{1/2}$, is preserved, and leads to a $\mathbb{Z}_2$ invariant classifying the AFM topological phases. In the AFTI, the $\mathbb{Z}_2$ parameters of the undoubled cell, $\gamma_0$ and $\gamma_\pi$, which correspond to the $k_z = 0,\pi$ time reversal invariant planes, combine together when Brillouin zone is folded in half to make these $k_z = 0$ and $k_z=\pi$ planes coincide. The resulting new $\mathbb{Z}_2$ coefficient of the doubled cell is equal to $\gamma_0^d = \gamma_0 + \gamma_\pi = s$, which can take on a value of 0 or 1 depending on if the system is a trivial ($s=0$) or topologically protected ($=1$) phase. This type of $\Theta T_{1/2}$-symmetry protected AFTI phase was recently confirmed in MnBi$_2$Te$_4$ \cite{mnbi2te4}, but a realization of the originally proposed AFTI arising from AFM ordering on a face centered cubic (FCC) sublattice has yet to be discovered.

The series of first-row transition metal dichalcogenides with the pyrite structure display a range of electronic phases depending on the electron filling. The crystal field environment of the tilted sulfur octahedra around each transition metal atom results in the splitting of the $3d$ manifold into a lower $t_{2g}$ band and an upper $e_g$ band. ZnS$_2$, which has a $3d^{10}$ electronic configuration is a paramagnetic insulator \cite{zns2}. Moving down the series, CuS$_2$ ($3d^9$) is a superconductor\cite{cus2}, NiS$_2$ ($3d^8$) is an antiferromagnetic Mott insulator \cite{nis2-mit, nis2-mit2, nis2-mit3}, CoS$_2$ ($3d^7$) is a ferromagnetic Weyl metal \cite{cos2,cos2-weyl}, and FeS$_2$ ($3d^6$) becomes a small gap semiconductor due to a completely empty $e_g$ band \cite{fes2,fes2-ferro}.

The $3d^5$ electron configuration of MnS$_2$ results in an interesting deviation from the expected trend in material properties determined by the electronic filling in this series. The increased stability of a half-occupied $3d$ orbital leads to a competition between high-spin and low-spin states. Energetic proximity of these states leads to a pressure driven magnetic transition which is accompanied by a structural transformation and volume collapse.

In this work, we show that the complex electronic structure of MnS$_2$ under pressure is further enriched by a series of topological phases. Using first-principles electronic structure calculations we show that the recently discovered pressure-induced metalization in this material occurs due to a crossing of the antibonding S$_2^{2-}$ $3p$-$\sigma^\ast$ and Mn $3d$-$e_g$ bands at the $\Gamma$ point. The spin-orbit coupling creates an small indirect gap in this crossing, which we show is in fact a Z2AFTM phase, by performing a parity analysis and direct calculations of the $\mathbb{Z}_2$ topological invariant. The AFM ordering of the face centered cubic (FCC) Mn sublattice, thus makes this a realization of the original proposal of the 3D AFTI protected by $\Theta T_{1/2}$ symmetry \cite{moore-afti}. Further compression of the lattice results in a metallic Weyl phase \cite{pyrochore-iridates}. Both topological phases would exhibit an anomalous Hall effect; semi-quantized for the Weyl phase, and quantized for the Z2AFTM.

\section{II. Structure and Magnetism.}

MnS$_2$ crystallizes in the cubic $Pa\bar{3}$ space group (\myhash 205) at ambient pressure (Figure 1a). The sulfur atoms in the structure form covalently bonded dimers that behave effectively as single S$_2^{2-}$ units linking together the tilted MnS$_6$ octahedral clusters. The FCC sub-lattice of Mn moments arranges in the AFM-III type magnetic ordering below 48.2 K, doubling the cubic unit cell \cite{mns2-discovery}. This magnetic transition temperature increases with pressure, reaching 76K at 5.3 GPa \cite{mns2-jetp}.

\begin{figure}[t]
    %\captionsetup{justification=raggedright}
    \includegraphics[width=1.05\columnwidth]{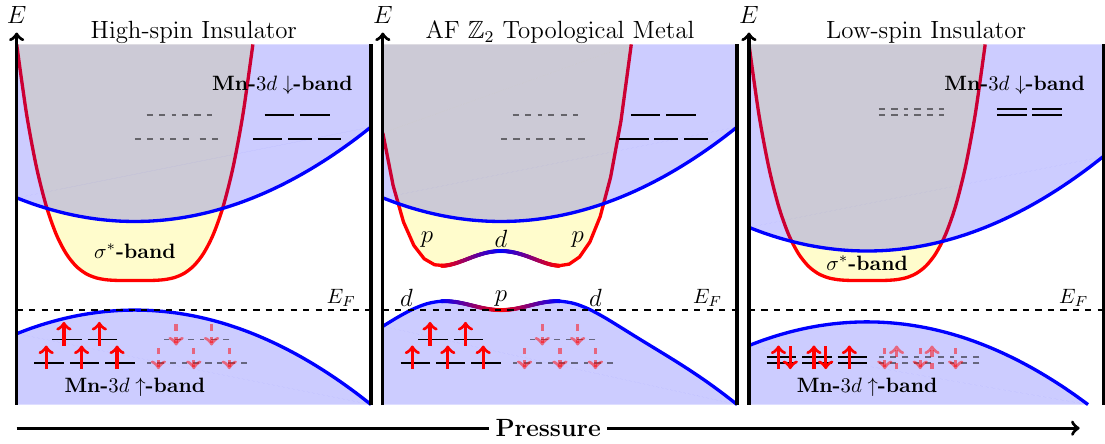}
    \caption{Schematic of the band-structure of MnS$_2$ at ambient pressure (left), near the insulator-to-metal transition (middle,) and high pressure (right). Dashed and solid lines/arrows correspond to states/electrons on opposite spin Mn sites.}
    \label{cartoon}
\end{figure}

At ambient pressure, the system is insulating with the d-band of Mn split into spin-up and spin-down, the lower of which is fully occupied, yielding an $S=5/2$ high spin state. The conduction band also has contributions from the S$_2^{2-}$ dimers, where the p-orbitals on S form an antibonded orbital $3p$-$\sigma^\ast = (p-p^\prime)/\sqrt{2}$. Upon applying pressure, the crystal field splitting between the $t_{2g}$ and $e_g$ bands increases (Figure \ref{cartoon}), resulting in all of the electrons occupying the $t_{2g}$ band, forming an $S=1/2$ low-spin state. The transition into the low-spin state is accompanied by a structural transition to the arsenopyrite structure at $\sim$11GPa, and a colossal volume collapse of 22\% -- the largest reversible volume collapse currently known\cite{mns2-vol-collapse}. The high-pressure structure with the low-spin state persists under decompression up to 6GPa before reverting back to the nominal pyrite structure. It was recently shown that the electronic transformation to the low-spin state under increasing pressure precedes and may in fact drive the structural transformation \cite{mns2-XANES}.

Both the high-spin and low-spin phases show characteristics of Mott insulators \cite{mns2-corr, mns2-spin-corr}, though more recent work has concluded that the material is in fact a charge transfer insulator \cite{mns2-vol-collapse}. Pressure dependent conductivity measurements revealed a metal-insulator-transition (MIT), marked by an eight order-of-magnitude increase in conductivity just before the spin-state crossover \cite{mns2-vol-collapse}.

To model the electronic structure of MnS$_2$ across the magnetic transition under pressure, the compressed structures were first relaxed using the full potential linearized augmented plane wave (FLAPW) method as implemented in WIEN2K \cite{wien2k}, starting from the experimental pyrite structure. The structural relaxations were performed using a LDA+SO+$U$ approach, with spin-orbit coupling, and Coulomb interactions were taken into account in the self-interaction correction (SIC) approximation \cite{Anisimov1993_LDAU}.

To ensure that our conclusions are not dependent on the parameters of the LDA+SO+$U$ method, we perform our calculations with both $U=4$eV and $U=5$eV, and a value of $J=1$eV for the Hund's coupling. For the experimental lattice parameters at ambient pressure, our calculation reproduces the experimentally observed AFM Type-III ordering of the magnetic moments, consistent with prior simulations of this material \cite{mns2-XANES}. The computed magnetic moment on the Mn atoms is 4.42$\mu_B$ for $U=4$eV, while for $U=5$eV it increases slightly to 4.47$\mu_B$. These values are consistent with a high-spin $S = 5/2$ state. 

\begin{table}[b]
    \begin{tabular}{ c | c c | c c }
        \hline
        \hline
        Compression & \multicolumn{2}{c|}{ $U = 4.0$eV} & \multicolumn{2}{c}{ $U = 5.0$eV} \\
        $1-V/V_0$ & $M(\mu_B)$ & phase & $M(\mu_B)$ & phase  \\\hline
        0\%& 4.42 & NI   & 4.47 & NI\\
        25\%& 3.80 & NI   & 4.02 & NI\\
        26\%& 3.74 & NI   & 3.97 & NI \\
        27\%& 3.51 & Z2AFTM &	3.92 & NI \\
        28\%& 2.98 & WM	  & 3.84 & NI \\
        29\%& 1.24 & NI	  & 3.73 & Z2AFTM \\
        30\%& 1.22 & NI	  & 2.59 & WM \\
        \hline
        \hline
    \end{tabular}
    \caption{Mn magnetic moments for different compressions of the MnS$_2$ pyrite structure, for varying values of Hubbard $U$. The phase at each pressure is indicated as normal insulator (NI), weyl metal (WM), or anti-ferromagnetic $\mathbb{Z}_2$ topological metal (Z2AFTM). }
    \label{moments}
\end{table}

Both the $U=4$eV and $U=5$eV cases are semiconductors at ambient pressure, with relatively large gaps of 1.23eV and 1.40eV respectively. As the pressure is increased, the highest occupied Mn $3d-e_g$ bands are pushed upward in energy due to increasing crystal field interactions, bringing them closer to the unoccupied S$_2 \sigma^\ast$ band just above $E_F$, leading to a narrowing of the gap. Under sufficiently high compression of the lattice the valence and conduction bands cross, leading to a metallic state, before the material reverts back to an insulating state at even higher pressures. The compression also suppresses the magnetic moment, which decreases rapidly after the onset of pressure induced metalization, and ultimately collapses into a low-spin state simultaneously with the reappearance of the electronic gap.

In order to more closely examine these various transitions under pressure, a wide range of lattice compressions was scanned to determine the region of interest for the metalization and suppression of the magnetic moment. The band crossing at $E_F$ and subsequent transformation into the low-spin state was found to occur within a narrow range of 25\% to 30\% volume compression. The relaxed structures between 25\% and 30\% in increments of 1\% were used to perform calculations using LDA+SO+$U$ within the full-potential muffin-tin orbital framework (FP-LMTO), to validate the FLAPW results and compute topological properties.

\begin{figure}[t]
    %\captionsetup{justification=raggedright}
    \includegraphics[width=0.95\columnwidth]{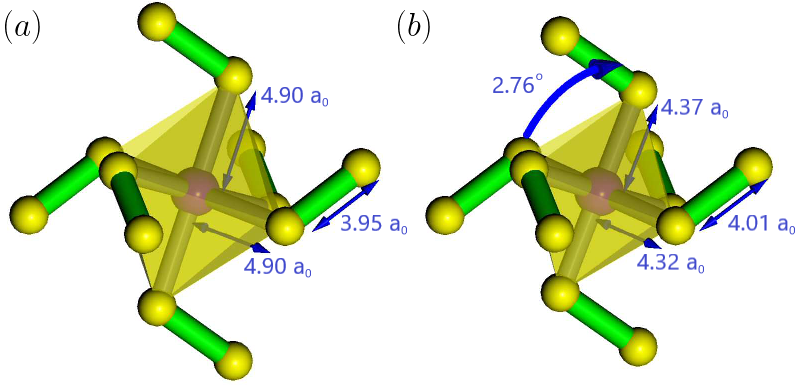}
    \caption{Structure of the Mn(S$_2$)$_6$ octahedra for a), the ideal experimental uncompressed structure and b), the computed relaxed structure at 27\% compression. Bond lengths in Bohr ($a_0$) are indicated, as well as the rotation of the main octahedron in degrees. 
    }
    \label{struct_change}
\end{figure}

Under pressure, the relaxed structure of MnS$_2$ undergoes a number of changes as compared to the ideal uncompressed pyrite structure. The magnetic order breaks the cubic symmetry of the structure, reducing it to the orthorhombic magnetic space group P$_\text{b}$ca2$_1$. This means that during the structural relaxation, the atoms are free to move as long as they respect the orthorhombic symmetry, so the final atomic positions of the structure may break the cubic $Pa\bar{3}$ symmetry as well. Fig \ref{struct_change} shows that as the structure is compressed, the octahedra rotate slightly by $2.76^\circ$, and are also significantly compressed. On the other hand, the sulfur dimers expand slightly despite the overall structural compression, and a Jahn-Teller-like distortion develops in the octahedra. All of these structural changes work in tandem to drive a band inversion under high pressure. The slightly differing lengths of the octahedral bonds break the degeneracy of both the Mn $3d$-$e_g$ bands below $E_F$, and the S$_2^{2-}$ $3p$-$\sigma^\ast$ bands above $E_F$, while the shortening of the bonds results in a greater overlap of the Mn and S$_2^{2-}$ centered orbitals, reducing the size of the band gap. Furthermore, the lengthening of the S$_2^{2-}$ dimers results in a lower energy for the antibonding $3p$-$\sigma^\ast$ orbital, forming an electron-like band around the $\Gamma$ point which dips in energy below the rest of the manifold, and ultimately inverts with the hole-like Mn $3d$-$e_g$ band at higher pressures. To further understand how the band inversion and metalization is driven by the structural distortion, we compute the band structures at different compression levels while keeping the relative atomic positions fixed to those of the high-pressure end-state \cite{supplement}. The band inversion occurs much earlier, at compressions as low as 20\%, showing that the structural distortion in fact helps drive the band inversion. In experiments, the insulator-to-metal transition occurs at much lower pressures than predicted by theory, suggesting that this structural distortion may in fact already be present at lower pressures. 

The computed magnetic moments for the compressed structures across the magnetic transition are shown in Table \ref{moments}. For the calculation with $U=4$eV, the high spin state at 25\% compression has a magnetic moment that is reduced slightly from theoretical ideal value for the $S=5/2$ magnetism, yielding a value of 3.8$\mu_B$. With pressure the moment decreases to 3.5$\mu_B$ at 27\% compression, where the bands first cross. Compressing beyond this volume results in a rapid reduction of the Mn magnetic moment to 1.2$\mu_B$ at 30\% compression, which is consistent with $S=1/2$ magnetism. The calculations at $U=5$eV follow a similar trend, though the magnetic moments are slightly larger, and a higher compression is needed in order to drive the transition into $S=1/2$ magnetism. This change in the electronic structure is known to precede the collapse to the arsenopyrite structure \cite{mns2-vol-collapse, mns2-XANES}, thus at these higher pressures the ambient pyrite structure may no longer be the ground state.

\begin{figure}[t]
    %\captionsetup{justification=raggedright}
    \includegraphics[width=0.95\columnwidth]{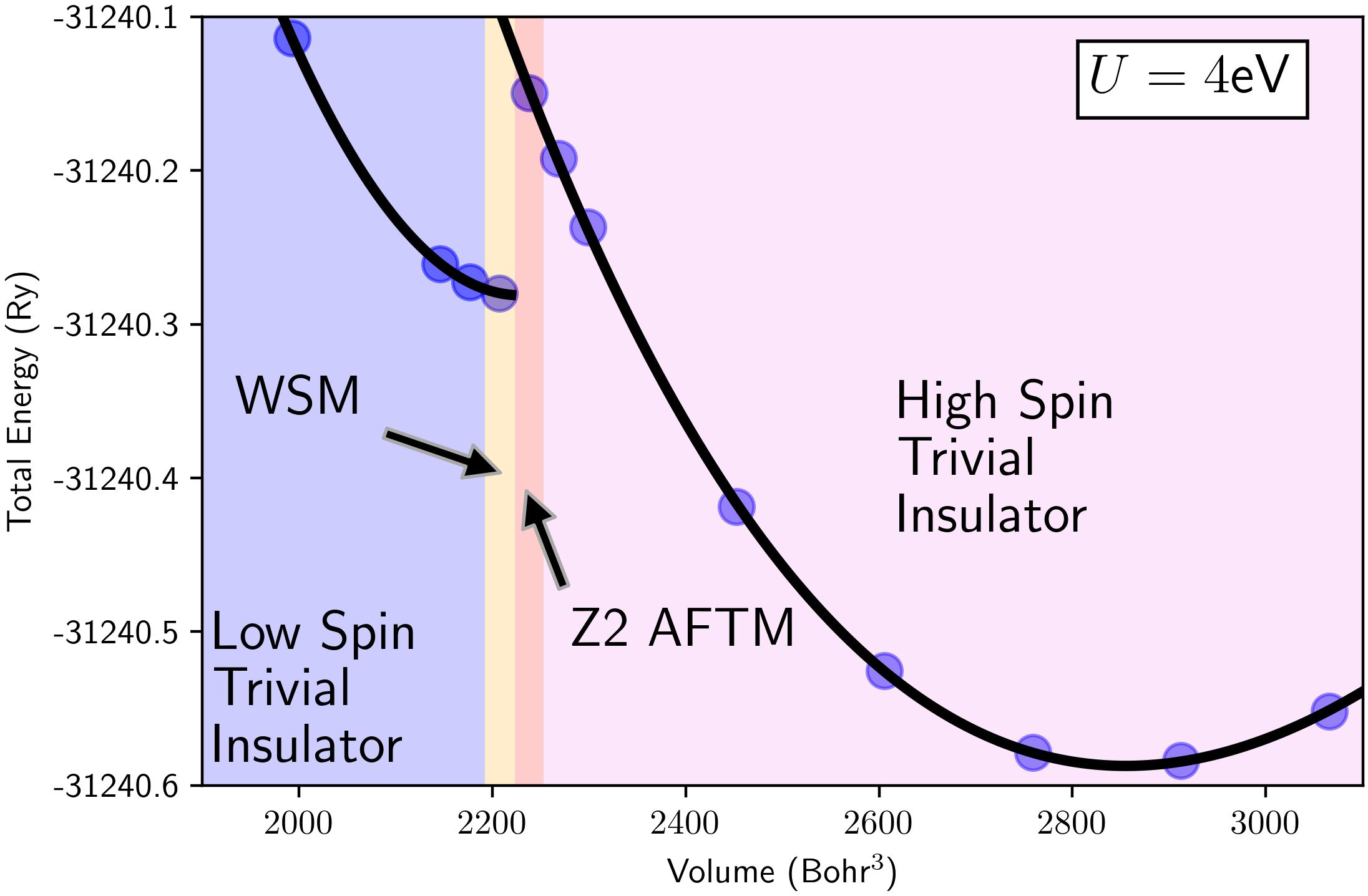}
    \caption{Total energy of MnS$_2$ computed for different unit cell volumes for $U=4$eV. The black lines is the Birch-Murnaghan equation of state fitted to the high-spin phase structures. Shaded panels denote the various phases occuring at different pressures: blue -- low-spin trivial insulator, orange -- WSM, red -- AFTI, violet -- high-spin trivial insulator.
    }
    \label{eos}
\end{figure}

The calculations at varying pressures were fitted to the Birch-Murnaghan \cite{BM-eos} equation of state (Figure \ref{eos}). The theoretical equation of state fits the computed energies of MnS$_2$ very well up to the point of metalization, but deviates significantly for higher compressions, consistent with a transition to a distinct electronic phase. The fit also allows us to extract the external pressure needed to stabilize each lattice volume. For the $U=4$eV calculation, an expansion pressure of -4.8 GPa is necessary, since computed equilibrium volume is in fact 2.3\% smaller than the experimental value. On the other hand for $U=5$eV, compression with a pressure of 4.0 GPa is needed, as the equilibrium volume is 1.7\% larger. The insulator to metal transition takes place at a pressure of 23.8 GPa (27\% compression) for $U=4$eV, and 45.8 GPa (29\% compression) for $U=5$eV (See Supplementary Material at \cite{supplement}), compared to the experimental value of 11 GPa. Despite the quantitative disagreement with experiment, these simulations correctly reproduce the insulator-to-metal-to-insulator transition and concomitant suppression of the magnetic moment, and are consistent with prior calculations which placed the metalization around $\sim20$ GPa \cite{mns2-XANES}.

The subsequent discussion will focus on the calculations with $U=4$eV due to the better quantitative agreement with the experimentally measured insulator to metal transition, though it should be noted that the results do not depend on the simulation parameters, with higher values of $U$ simply requiring higher pressures to reproduce the same phases.

\section{III. Topology}

\begin{figure}[hb]
	%\captionsetup{justification=raggedright}
	\includegraphics[width=0.95\columnwidth]{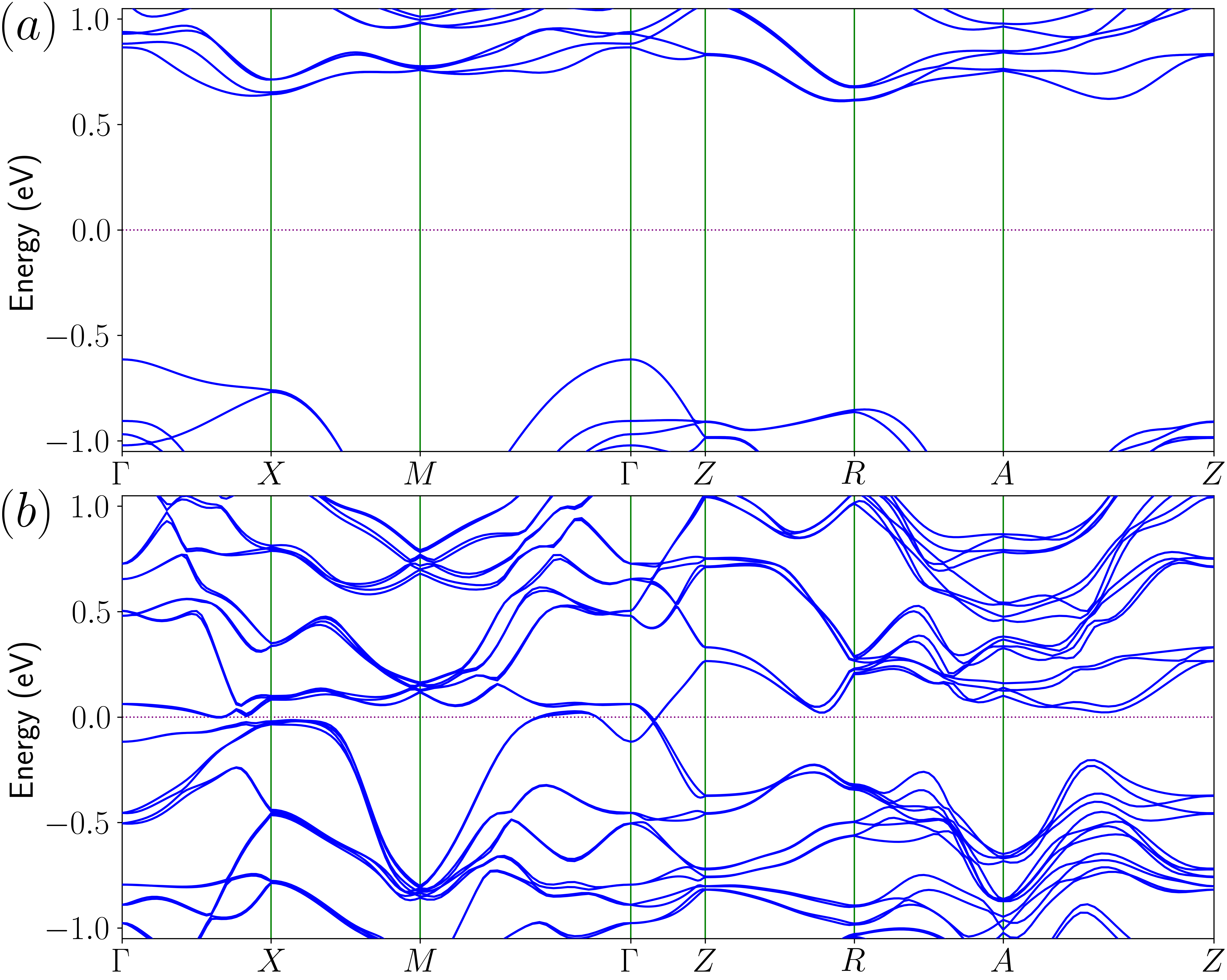}
	\caption{Band structures of MnS$_2$ for (a) no compression and (b) 27\% compression. In (b) the bands invert around the $\Gamma$ point, and there is a gap throughout the Brillouin zone. }
	\label{bands}
\end{figure}

We will now discuss the topological properties of MnS$_2$. The parity criterion \cite{fukane, Turner2012} for identifying topologically nontrivial phases is valid in systems possessing inversion symmetries, as is the case for MnS$_2$, and is therefore useful for understanding the origin of the topology. Ignoring the magnetism, at the time-reversal invariant $\Gamma$ point the odd parity S$_2^{2-}$ $3p$-$\sigma^\ast$ band crosses $E_f$, inverting with the lower-lying even parity Mn $3d$-$e_g$ band (Fig. \ref{bands}). This band inversion is a strong indication of the presence of topological feature. 
%This flips the sign at a single time-reversal invariant momentum point, and hence inverts the sign of the overall parity criterion $(-1)^{\nu_0} = \prod^8_{i=1} \delta_i$, implying a strong topological insulator phase with $\nu_0 = 1$.

We now turn to the computational treatment of MnS$_2$, focusing on the 25\%, 26\%, and 27\% compressed structures. To verify the topological properties, special care must be taken to consider other possibilities that may arise. When either time reversal or inversion symmetries are broken, topological Weyl points \cite{pyrochore-iridates} may arise in the electronic band structure. These Weyl points may be located at arbitrary points in the Brillouin zone, and might exist despite the apparent gap in the band structure when plotting along high symmetry lines. To check for the presence of Weyl points we use an approach which locates sources and sinks of Berry curvature \cite{mmm}. This procedure computes an integer link field using the FP-LMTO electronic structure, eliminating the need for projecting onto Wannier orbitals. The search is carried out on a $20\times 20\times 10$ $\bm{k}$-point grid, and yields no topological point nodes for the 25-27\% compression range.

%Having excluded the possibility of a Weyl semimetal phase in the 25-27\% compression range, confirming an insulating state, we now determine whether the gap is topologically nontrivial. 
We now discuss the topological nature of the band gap in MnS$_2$ across the 25-27\% compression range. In MnS$_2$ the magnetic ordering breaks time-reversal symmetry, doubling the unit cell. 
%
%In effect, for the original unit cell, both time reversal symmetry $\Theta$ and half-translation symmetry $T_{1/2}$ are broken, but their combination $S=\Theta T_{1/2}$ is preserved. 
%
In effect, both time reversal symmetry $\Theta$ and half-translation symmetry $T_{1/2}$ are broken.
%
% Below removed per Xiangang's request
%For this type of symmetry, an explicit mapping exists between the $\mathbb{Z}_2$ invariants of the original cell and those of the doubled cell \cite{moore-afti}. 
%In the original cell, there are six planes satisfying time-reversal symmetry, $k_x = 0,\pi$, $k_y = 0,\pi$, and $k_z = 0,\pi$, with the respective $\mathbb{Z}_2$ invariants $\alpha_0, \alpha_\pi, \beta_0, \beta_\pi, \gamma_0$, and $\gamma_\pi$ taking on values of either 0 or 1. Their sum $s = \alpha_0 + \alpha_\pi = \beta_0 + \beta_\pi = \gamma_0 + \gamma_\pi$ (mod 2), constitutes a strong topological invariant, with $s=1$ corresponding to the topologically nontrivial state. When the unit cell is doubled along the $\bm{a}_z$ direction, the new lattice parameter becomes $\bm{a}^d_z = 2\bm{a}_z$, while the Brillouin zone is folded in half with the $\bm{k}_z = 0$ and $\bm{k}_z = \pi$ planes folding into each other to form the new $\bm{k}^d_z = 0$ plane. This results in the new invariant $\gamma_0^d = \gamma_0 + \gamma_\pi = s$. 
%replaced with this:
In particular, the $\mathbb{Z}_2$ invariant of the $\bm{k}^d_z = 0$ plane in the BZ of the doubled cell, is exactly equal to the strong topological invariant, $\gamma_0^d = \gamma_0 + \gamma_\pi = s$, with $s=1$ corresponding to the topologically nontrivial state.
Thus computing the new $\mathbb{Z}_2$ invariant $\gamma_0^d$ thus directly determines whether the material constitutes a topologically nontrivial phase as described by $s$.

The $\mathbb{Z}_2$ invariants for the structures in the 25-27\% compression range are computed using the link variable method \cite{fukui,link-fplapw}. This method is known to be quite stable, reproducing the integer topological invariants even for relatively coarse $\bm{k}$-point grids. To compute the invariants, we use 20 divisions of the Brillouin Zone to define the grid. The procedure uncovers a topologically trivial $(\gamma_0^d = s = 0)$ phase at 25\% compression, and topologically non-trivial $(\gamma_0^d = s = 1)$ phase at 26\% - 27\% compression. It should be noted that while this confirms a topological insulator phase, at 27\% compression there are bands crossing $E_f$, and at 26\% the gap is $\sim 10$ meV, meaning the system is effectively metallic. Convergence of the  $\mathbb{Z}_2$ invariant calculation was checked for $\bm{k}$-point grids up to $50 \times 50$, confirming the existence of the topological transition.

The essential features of the Z2AFTM phase in MnS$_2$ can be captured using a simple tight binding model using a basis of a Mn-$3d$ orbital and the $\sigma^*$ orbital of the S$_2$ dimers, both of which lie on FCC sublattices, that taken together form a rock-salt type structure. The effective Hamiltonian can be written as 
\begin{equation}
    \mathcal{H} = \mathcal{H}_0 + \mathcal{H}_{d\text{-}\sigma^*}  + \mathcal{H}_{\text{so}} + \mathcal{H}_{\text{mag}}.
\end{equation}
The first term represents the onsite energies of the S$_2$-$\sigma^*$ orbitals ($p^\dagger_{i\alpha}$, $p_{i\alpha}$), and Mn-$3d$ orbitals ($d^\dagger_{i\alpha}$, $d_{i\alpha}$),
\begin{equation}
    \mathcal{H}_0 = \epsilon_{\sigma^*} \sum_{i\alpha} p^{\dagger}_{i\alpha} p_{i\alpha}^{\phantom{\dagger}} + \epsilon_{d} \sum_{i\alpha} d^{\dagger}_{i\alpha} d_{i\alpha}^{\phantom{\dagger}},
\end{equation}
where $i$ and $\alpha$ represent the site and spin indices respectively. The second term is comprised of the nearest neighbor and next-nearest neighbor hoppings between orbitals,
\begin{align}
\mathcal{H}_{d\text{-}\sigma^*} &= \sum_{ij\alpha} \left[ t_{\sigma^*} p^{\dagger}_{i\alpha} p_{j\alpha}^{\phantom{\dagger}} + t_{d} d^{\dagger}_{i\alpha} d_{j\alpha}^{\phantom{\dagger}} \right. \nonumber\\ 
&+  \left. t_{d\sigma^*} \left( d^{\dagger}_{i\alpha} p_{j\alpha}^{\phantom{\dagger}}+ p^{\dagger}_{i\alpha} d_{j\alpha}^{\phantom{\dagger}} \right) \right].
\end{align}
The spin orbit couples between the non-magnetic orbitals,
\begin{equation}
\mathcal{H}_{\text{so}} = i t_{\text{so}} \sum_{ij\alpha\alpha^\prime}  p^{\dagger}_{i\alpha} \bm{\sigma}\cdot\left( \bm{d}_{ij}^{1} \times \bm{d}_{ij}^{2} \right) p_{j\alpha^\prime}^{\phantom{\dagger}},
\end{equation}
where $\bm{\sigma} = ( \sigma_x , \sigma_y , \sigma_z)$  is the vector of Pauli matrices, and $\bm{d}_{ij}^{1}, \bm{d}_{ij}^{2}$ are the two nearest neighbor bond vectors that are traversed between the second-nearest neighbor sites $i$ and $j$. In particular, this term is only non-zero when the two intermediate Mn sites have equal magnetization, otherwise the cross products $\left( \bm{d}_{ij}^{1} \times \bm{d}_{ij}^{2} \right)$, will cancel \cite{moore-afti}. 
Finally, antiferromagnetic order is introduced through the term $\mathcal{H}_{\text{mag}} = \lambda s_i \sigma_z$, where $s_i = +1$ or $-1$ for spin up or spin down Mn atoms, and $\sigma_z$ is the Pauli matrix acting on the electron spin degree of freedom. 

In order to qualtatively reproduce the $\mathbb{Z}_2$ topological phase in MnS$_2$, we take the model paramaters to be $\epsilon_{d} = -2.2$, $\epsilon_{\sigma^*} = 1.9$, $t_{\sigma^*} = -0.15$, $t_{d} = 0.2$, $t_{d\sigma^*} = 0.5$, and magnetic splitting $\lambda = 0.2$. As the spin-orbit coupling $0<t_{\text{so}}<0.3$ is increased, the valence band maximum and conduction band minimum invert at the gamma point. It should be noted that in the real material, the spin-orbit coupling will be much smaller than the hopping $t_{\text{so}} \ll t_{d\sigma^*}$. In the model, the same physics can be recovered with a much smaller $t_{\text{so}}$ if the gap is decreased by lowering  $\epsilon_{\sigma^*}$, however, we choose a larger value in order to more clearly visualize the surface states. A 10-layer slab calculation reveals that when the bands are inverted by spin-orbit coupling, a single Dirac cone appears on the $z$-surface, protected by the combined time-reversal and half-translation symmetries. The exact form of the Hamiltonian used is provided in the supplementary material \cite{supplement}, along with details on the slab calculations of the topological surface states.

\begin{table}[t]
    \begin{tabular}{ c c c}
        \hline
        \hline
        Position & $E-E_F$ (meV) & Chirality   \\\hline
        (0.2126, 0.3325, 0.0413) & -287 meV & $-1$ \\
        (0.1896, 0.4140, 0.2747) & -283 meV & $-1$ \\
        (0.2891, 0.1111, 0.4618) &  -54 meV & $-1$ \\
        (0.3487, 0.1251, 0.0609) & -131 meV & $-1$ \\
        (0.2092, 0.4024, 0.2420) &  -61 meV & $-1$ \\
        (0.3917, 0.0507, 0.0340) & +129 meV & $+1$ \\
        \hline
        \hline
    \end{tabular}
    \caption{The $k$-space locations in units of $(2\pi/a, 2\pi/a, 2\pi/a)$, energy relative to $E_F$, and chirality of the symmetry-inequivalent Weyl points in the Weyl metal phase of MnS$_2$.}
    \label{weyls}
\end{table}

Continuing to 28\% compression of the lattice results in a metallic Weyl phase, for which the previously described Weyl point search procedure \cite{mmm} yields a number of symmetry protected Weyl points in the Brillouin zone between the two bands crossing at $E_F$. The cubic lattice symmetry of MnS$_2$ is broken by AFM order in the $z$ direction, so the Weyl points generate in symmetry-related sets of eight at the positions given in Table \ref{weyls}. Each listed Weyl has symmetry-related members found at $(\pm k_x, \pm k_y, \pm k_z)$, with Weyl points related by a reflection having opposite chirality. 

Both the Weyl metal and Z2AFTM phases should exhibit an anomalous Hall effect (AHE) \cite{burkov,moore-afti} that can be observed experimentally to detect the onset of these topological phases under pressure. In the Z2AFTM phase, the surface of MnS$_2$ would have a 2D half-quantized AHE $\sigma^{2D}_{xy} = e^2/(2h)$. In the Weyl metal phase, this exact quantization would break, due to the presence of Weyl points. Instead, each pair of Weyls would have a contribution \cite{qh_weyl} $\sigma^{3D}_{xy} = e^2k_W/(2\pi h)$ to the 3D AHE, where $k_z$ is the separation in $k$-space between the Weyl points comprising the pair.

\section{IV. Conclusion}

Our prediction of Z2AFTM and Weyl metal phases within the spin-state crossover transition in MnS$_2$, along with the recent confirmation of a Weyl semimetal phase in CoS$_2$, has now cemented topology as yet another functional property of pyrite-type transition metal disulfides which already included strongly correlated Mott physics, superconductivity, and gate-induced magnetism. These results further reinforce the status of these transition metal disulfides as multifunctional materials for experimentally realizing a diverse range of exotic electronic properties.

The Z2AFTM phase in MnS$_2$ and Weyl semimetal phase in CoS$_2$ also motivate the search for topological phases in other pyrite-type TMDCs, which may be magnetic at higher temperatures or wider gaps due to stronger spin-orbit coupling from heavier elements. One such material is MnTe$_2$ \cite{mnte2}, which orders antiferromagnetically at 86.5K, and also undergoes a pressure induced insulator-metal transition. Another possibility is the Mott insulator NiS$_2$, which has AFM order below ~40K \cite{nis2-mit,nis2-mit2,nis2-mit3}, and can be made metallic either by applying moderate pressure or doping with Se at ambient pressure. If the topological phases were to exist, these materials would have much more suitable properties for applications such as AFM spintronics.

\section{\textbf{Acknowledgements}}
\begin{acknowledgments}
%VI was supported by the Molecular Foundry, a DOE Office of Science User Facility supported by the Office of Science of the U.S. Department of Energy under Contract No. DE-AC02-05CH11231.
\end{acknowledgments} 

\bibliography{references}

\end{document}